\documentclass[12pt,english,nofootinbib,preprint]{revtex4-1}
\usepackage{hyperref}
\usepackage{setspace}
\usepackage{mathptmx}
\usepackage{graphicx}

\usepackage[T1]{fontenc}
\usepackage[latin9]{inputenc}
\usepackage[letterpaper]{geometry}
\usepackage{booktabs}
\usepackage{amsmath}
\usepackage{amssymb}
\usepackage{cancel}
\usepackage{babel}

\begin{document}

\title{Critical Droplets and Replica Symmetry Breaking}
\begin{abstract}
We show that the notion of critical droplets is central to an understanding of the nature of ground states in the Edwards-Anderson Ising model of a spin glass in arbitrary
dimension. Given a specific ground state, suppose the coupling value for a given edge is varied with all other couplings held fixed.  Beyond some specific value of the coupling,
a droplet will flip leading to a new ground state; we refer to this as the critical droplet for that edge and ground state. We show that the distribution of sizes and energies over all
edges for a specific ground state can be used to determine which of the leading scenarios for the spin glass phase is correct. In particular, the existence of low-energy interfaces
between incongruent ground states as predicted by replica symmetry breaking is equivalent to the presence of critical droplets whose boundaries comprise a positive fraction of edges in the infinite lattice.
\end{abstract}
\author{C.M.~Newman}
\affiliation{ Courant Institute of Mathematical Sciences, New York University, New York, NY 10012 USA }
\affiliation{NYU-ECNU Institute of Mathematical Sciences at NYU Shanghai, 3663 Zhongshan Road North, Shanghai, 200062, China}
\author{D.L.~Stein}
\affiliation{Department of Physics and Courant Institute of Mathematical Sciences, New York University, New York, NY 10012 USA}
\affiliation{NYU-ECNU Institutes of Physics and Mathematical Sciences at NYU Shanghai, 3663 Zhongshan Road North, Shanghai, 200062, China}
\affiliation{Santa Fe Institute, 1399 Hyde Park Rd., Santa Fe, NM USA 87501}

\maketitle

\section{Introduction}
\label{sec:intro}

The nature of the low-temperature phase of the Edwards-Anderson Hamiltonian~\cite{EA75} in finite dimension
\begin{equation}
\label{eq:EA}
{\cal H}_J=-\sum_{<x,y>} J_{xy} \sigma_x\sigma_y 
\end{equation}
remains unresolved.  Here $\sigma_x=\pm 1$ is the Ising spin at site $x$ and $\langle x,y\rangle$ denotes a nearest-neighbor edge in the edge set $\mathbb{E}^d$ of the $d$-dimensional cubic lattice~$\mathbb{Z}^d$. The couplings $J_{xy}$ are taken to be independent, identically distributed continuous random variables chosen from a distribution $\nu(dJ_{xy})$, with random variable $J_{xy}$  assigned to the edge $\langle x,y\rangle$.  Our requirements on $\nu$ is that it be supported on the entire real line, distributed symmetrically about zero, and has finite variance; e.g., a Gaussian with mean zero and variance one.  We denote by $J$ a particular realization of the couplings.

There are at present four scenarios for the spin glass phase that are consistent both with numerical results and, as far as is currently known, mathematically consistent: replica symmetry breaking (RSB)~\cite{Parisi79,Parisi83,MPSTV84a,MPSTV84b,MPV87,MPRR96,MPRRZ00,NS02,NS03b,Read14,NRS23b}, droplet-scaling~~\cite{Mac84,BM85,BM87,FH86,FH88b}, trivial-nontrivial spin overlap~(TNT)~\cite{MP00,PY00}, and chaotic pairs~\cite{NS96c,NS97,NSBerlin,NS03b}.   One of the central open questions in spin glass theory concerns which (if any) of these pictures are correct, and for which dimensions and temperatures.

The differences among the four pictures at positive temperature are described elsewhere~\cite{NS03a,NS22,NRS23b}; here we are concerned with their different predictions at zero temperature, i.e., for the ground state structure of the EA~Hamiltonian.  Of the four, two (RSB and chaotic pairs) predict the existence of many ground states, and the other two (scaling-droplet and TNT) imply the existence of only a single pair~\cite{FH88b,NS01c,ANS19}.  While important, these differences are less fundamental than the nature of the {\it interfaces\/} that separate their ground states from their lowest-lying long-wavelength excitations. The presence or absence of multiplicity of ground states follows as a consequence of the nature of these excitations.

In this paper we focus on the nature of low-energy large-lengthscale excitations above the ground state and how they relate to ground state stability, with a view toward distinguishing different predictions of the four pictures.  Aside from elucidating the different (and potentially testable) predictions of these pictures, determining the stability properties of the ground state is crucial in determining the low-temperature properties of the spin glass phase, including central questions such as multiplicity of pure states and the presence or absence of an AT line~\cite{AT78}.  We begin by defining the objects of study.

Take a finite volume~$\Lambda_L$, chosen to be a cube of side~$L$ centered at the origin. A finite-volume ground state~$\sigma_L$ is the lowest-energy spin configuration in~$\Lambda_L$ subject to a specified boundary condition. An infinite-volume ground state~$\sigma$ is a spin configuration on all of $\mathbb{Z}^d$ defined by the condition that its energy cannot be lowered by flipping any {\it finite\/} subset of spins.  (Of course, $\sigma$ is always defined with respect to a specific~$J$, but we suppress its dependence for notational convenience.)  The condition for $\sigma$ to be a ground state is then
\begin{equation}
\label{eq:gs}
E_{S}=\sum_{\langle x,y\rangle\in{S}}J_{xy}\sigma_x\sigma_y\ >0 
\end{equation}
where $S$ is any closed surface (or contour in two dimensions) in the dual lattice. The surface~$S$ encloses a connected set of spins (a ``droplet''), and $\langle x,y\rangle\in{S}$ is the set of edges connecting spins in the interior of~$S$ to spins outside $S$.   The inequality in~(\ref{eq:gs}) is strict since, by the continuity of $\nu(dJ_{xy})$, there is zero probability of any closed surface having exactly zero energy in $\sigma$. The condition~(\ref{eq:gs}) must also hold for finite-volume ground states for any closed surface completely inside $\Lambda_L$. It is then not hard to show that an alternative (and equivalent) definition, which we also sometimes use, is that an infinite-volume ground state is any convergent limit of an infinite sequence of finite-volume ground states. Given the spin-flip symmetry of the Hamiltonian, a ground state, whether finite- or infinite-volume, generated by a spin-symmetric boundary condition, such as free or periodic, will appear as one part of a globally spin-reversed pair; we therefore refer generally to ground state pairs (GSP's) rather than individual ground states.

\section{Interfaces and Critical Droplets}
\label{subsec:interfaces}

An interface between two infinite-volume spin configurations $\alpha$ and $\beta$ comprises the set of edges whose associated couplings are satisfied in $\alpha$ and unsatisfied in $\beta$, or vice-versa; they separate regions in which the spins in $\alpha$ agree with those in $\beta$ from regions in which their spins disagree. An interface may consist of a single connected component or multiple disjoint ones, but (again using the continuity of the coupling distribution) if $\alpha$ and $\beta$ are ground states any such connected component must be infinite in extent. 

Interfaces can be characterized by their geometry and energy. They can be either ``space-filling'', meaning they comprise a positive density of all edges in~$\mathbb{E}^d$, or zero-density, in which the dimensionality of the interface is strictly less than the dimension $d$.  Ground states are called {\it incongruent\/} if they differ by a space-filling interface~\cite{HF87,FH87}. 

Interfaces can also differ by how their energies scale with volume. The energy might diverge (though not monotonically) as one examines interfaces contained within increasingly larger volumes, or it might remain~$O(1)$ independent of the volume considered. We will denote the former a ``high-energy interface'' and the latter a ``low-energy interface''.

An excitation above the ground state is any spin configuration obtained by overturning one or more spins in the ground state (while leaving an infinite subset of spins in the original ground state intact), and so an interface is the boundary of an excitation.  We will be primarily interested in excitations consisting of overturning droplets of large, or possibly infinite, size; because an interface is the boundary of such an excitation, the energy of the excitation is simply twice the interface energy.  An excitation above a ground state may itself be a new ground state (this would require the excitation to involve overturning an infinite number of spins such that~(\ref{eq:gs}) remains satisfied). Indeed, as proved elsewhere~\cite{NS02}, an excitation having a space-filling interface with the original ground state may generate a new ground state entirely. 

With this in mind we present the four low-temperature spin glass scenarios in the following table, which illustrates their various relationships (and clarifies why we consider these four pictures together):

\bigskip

\begin{center}
\begin{tabular}{ r|c|c| }
\multicolumn{1}{r}{}
 &  \multicolumn{1}{c}{Low-energy}
 & \multicolumn{1}{c}{High-energy} \\
\cline{2-3}
Space-filling&RSB&Chaotic pairs \\
\cline{2-3}
 Zero-density&TNT&Scaling-droplet\\
\cline{2-3}
\end{tabular}

\medskip

\begin{singlespace}
\small{{\bf Table~1.} The four scenarios described in the text for the low-temperature phase of the EA model, categorized in terms of interface geometry (rows) and energetics (columns). The column headings describe the energy scaling along the interface of the minimal large-lengthscale excitations above the ground state predicted by each. Adapted from Fig.~1 of~\cite{NS03a}.}
\end{singlespace}

\bigskip

\end{center}

As shown elsewhere~\cite{NS02}, the existence of space-filling interfaces in the first row scenarios (RSB and chaotic pairs) implies the presence of multiple GSP's, while droplet-scaling and TNT both predict a single ground state pair~\cite{HF87,FH87,NS01c,NS02,ANS19}.

{\bf Remark on Table~1.} The scaling-droplet picture predicts a broad distribution of (free) energies for a minimal energy compact droplet of diameter~$O(L)$, with a characteristic energy growing as $L^\theta$ with $\theta>0$ in dimensions where a low-temperature spin glass phase is present. The distribution is such that there exist droplets of~$O(1)$~energy on large lengthscales, but these appear with a probability falling off as $L^{-\theta}$ as $L\to\infty$. In contrast, both the RSB and TNT pictures require droplets with $O(1)$ energy to appear with positive probability bounded away from zero on all lengthscales. Thus the scaling-droplet scenario belongs in the second column of Table~1.

\medskip

We now turn to the concepts of flexibility and critical droplets, which were introduced in~\cite{NS2D00,NS2D01} and whose properties were described extensively in~\cite{NS22} (see also~\cite{ANS19,ANS21}), to which we refer the reader for a detailed description. Here we only summarize their main features. We first provide some definitions (all with respect to some fixed coupling realization~$J$):

\medskip

{\bf Definition 2.1~[Newman-Stein~{\cite{NS22}]}.} Consider the ground state pair $\sigma_L$ for the EA Hamiltonian~(\ref{eq:EA}) on a finite volume $\Lambda_L$ with boundary conditions chosen independently of $J$ (for specificity we always use periodic boundary conditions (PBC's) in this paper). Choose an edge $b_{xy}=\langle{x,y}\rangle$ with $x,y\in\Lambda_L$ and consider all closed surfaces in the dual edge lattice ${\mathbb E}_L^*$ which include the dual edge $b^*_{xy}$. By~(\ref{eq:gs}) and the continuity of the couplings, these all have distinct positive energies. There then exists a closed surface $\partial D(b_{xy},\sigma_L)$, passing through $b^*_{xy}$, of {\it least\/} energy in $\sigma_L$. We call $\partial D(b_{xy},\sigma_L)$ the {\it critical droplet boundary\/} of $b_{xy}$ in $\sigma_L$ and the set of spins $D(b_{xy},\sigma_L)$ enclosed by $\partial D(b_{xy},\sigma_L)$ the {\it critical droplet\/} of $b_{xy}$ in $\sigma_L$.

\medskip

{\bf Remark.} Critical droplets are defined with respect to edges rather than associated couplings to avoid confusion, given that we often vary the coupling value associated with specific edges, while the edges themselves are fixed, geometric objects.

\medskip

We define the energy~$E\bigl(D(b_{xy},\sigma_L)\bigr)$ of the critical droplet of $b_{xy}$ in $\sigma_L$ to be the energy of its boundary as given by~(\ref{eq:gs}):
\begin{equation}
\label{eq:cd}
E\bigl(D(b_{xy},\sigma_L)\bigr)=\sum_{<x,y>\in\partial D(b_{xy},\sigma_L)}J_{xy}\sigma_x\sigma_y\, .
\end{equation}

\medskip

{\bf Definition 2.3.  [Newman-Stein~{\cite{NS22}]}} The {\it critical value\/} of the coupling $J_{xy}$ associated with $b_{xy}$ in $\sigma_L$ is the value of $J_{xy}$ where $E\bigl(D(b_{xy},\sigma_L)\bigr)=0$, while all other couplings in $J$ are held fixed.

\medskip

We next define the {\it flexibility\/} $f(J_{xy},\sigma_L)$:
\medskip 

{\bf Definition 2.4.  [Newman-Stein~{\cite{NS22}]}} Let $J_{xy}$ be the value of the coupling assigned to the edge~$b_{xy}$ in coupling realization~$J$ and $J_c(b_{xy},\sigma_L)$ be the critical value of $b_{xy}$ in $\sigma_L$.  We define the flexibility $f(b_{xy},\sigma_L)$ of $b_{xy}$ in $\sigma_L$ to be $f(b_{xy},\sigma_L)=|J_{xy}-J_c(b_{xy})|$. 

\medskip

{\bf Remark.} The critical value $J_c$ of an edge $b_{xy}$ with coupling value $J_{xy}$ is determined by all couplings in $J$ {\it except\/} $J_{xy}$.  Because couplings are chosen independently from $\nu(dJ_{xy})$, it follows that the value $J_{xy}$ is {\it independent\/} of~$J_c$. Therefore, given the continuity of $\nu(dJ_{xy})$, there is zero probability in a ground state that any coupling has exactly zero flexibility.

\smallskip

It follows from the definitions above that
\begin{equation}
\label{eq:flex}
f(b_{xy},\sigma_L)=E\bigl(D(b_{xy},\sigma_L)\bigr)\, .
\end{equation}
Therefore couplings which share the same critical droplet have the same (strictly positive) flexibility. 

A rigorous definition of critical droplets and flexibilities within infinite-volume ground states requires use of the excitation metastate, whose definition and properties were presented in~\cite{NS2D00,NS2D01,ADNS10,ANS19}, to which we refer the interested reader.  Here we simply note that finite-volume critical droplets and their associated flexibilities converge with their properties preserved in the infinite-volume limit, for reasons presented in~\cite{NS22}.  This result would be trivial if all critical droplets in infinite-volume ground states were finite.  However, it could also be that critical droplets can be infinite in extent in one or more directions, in which case metastates can be used to define such unbounded critical droplets which enclose an infinite subset of spins: they are the infinite-volume limits of critical droplets in finite-volume ground states.

\section{Classification of critical droplets}
\label{sec:classify}

In~\cite{NS22} critical droplets in infinite-volume ground states were classified according to the size of their boundary~$\partial D(b_{xy},\sigma)$, which is the relevant factor in associating the presence of a given type of critical droplet with one of the pictures in Table~1.  We simplify the nomenclature used in that paper by focusing here on three different kinds of critical droplet. Let $\vert\partial D(b_{xy},\sigma)\vert$ denote the number of edges in the critical droplet boundary. A {\it finite\/} critical droplet is one in which $\vert\partial D(b_{xy},\sigma)\vert<\infty$; in two and more dimensions, this implies that the critical droplet~$D(b_{xy},\sigma)$ itself consists of a finite set of spins and thus can be completely contained within some finite volume.  (A $1D$ chain is the exception: here the critical droplet boundary of any edge consists of that edge alone, but the associated critical droplet consists of a semi-infinite chain of spins.)
If these are the only type of critical droplet present, then the distribution of their sizes becomes important in answering fundamental questions involving edge disorder chaos and ground state structure~\cite{ANS19}.  It is not hard to show that in any dimension, an EA ground state must contain at least a positive density of edges with finite critical droplets (while in $1D$ this is the case for {\it all\/} edges).

There are two kinds of critical droplets with $\vert\partial D(b_{xy},\sigma)\vert=\infty$. The first class includes those with infinite boundary $\partial D (b_{xy},\sigma)$ having a lower dimensionality than the space dimension~$d$; that is, the critical droplet boundary is infinite but zero-density~in $\mathbb{E}^d$.  We refer to these as ${\it zero-density\/}$ critical droplets~(ZDCD). (Of course a finite critical droplet boundary also has zero density in $\mathbb{E}^d$, but we reserve the term ``zero-density critical droplet'' to apply only to critical droplets with infinite boundary.) 

Finally, there is the possibility that there exist infinite critical droplets whose boundary has dimension~$d$, i.e., $\partial D(b_{xy},\sigma)$ comprises a positive density of edges in $\mathbb{E}^d$. We refer to these as {\it space-filling\/} critical droplets~(SFCD). These critical droplets have boundaries that pass within a distance~$O(1)$ of any site in~$\mathbb{Z}^d$, i.e., the closest distance from any site in $\mathbb{Z}^d$ to $\partial D(b_{xy},\sigma)$ is essentially independent of the location of the site.

Because our ground states are chosen from the zero-temperature PBC~metastate (denoted $\kappa_J$), we can adapt a result from~\cite{NS01c,NS24a,NS24b}, which is the following:

\medskip

{\bf Theorem 3.1.} Let $\sigma$ denote an infinite-volume spin configuration.  Then for almost every $(J,\sigma)$ pair at zero temperature (which restricts the set of $\sigma$'s to ground states corresponding to particular coupling realizations~$J$), and for any type of critical droplet (finite, zero-density, or positive-density), either a positive density of edges in $\sigma$ has a critical droplet of that type or else no edges do.

\medskip

The method of proof of this theorem is essentially identical to that used in~\cite{NS01c,NS24b} and so will be omitted here. 
The conclusion is that there is zero probability that a ground state~$\sigma$ chosen from~$\kappa_J$
has a (finite or infinite) set of edges with zero density in~$\mathbb{E}^d$ and having SFCD's (or finite critical droplets or ZDCD's).

\section{Critical droplets and replica symmetry breaking}
\label{sec:rsb}

In~\cite{NS22} it was shown that there is an intimate connection between critical droplets and the four pictures shown in Table~1.  However, the results obtained there were incomplete
for the most prominent of the four scenarios, namely replica symmetry breaking. In particular, it was proved there that the existence of SFCD's was a sufficient condition for some pairs of incongruent ground states to be separated by space-filling low-energy interfaces, hereafter referred to simply as ``RSB interfaces'' in accordance with Table~1. However, they were not shown to be necessary.  
The main goal of this paper is to complete the correspondence between critical droplets and spin glass scenarios by showing that the presence of SFCD's is not only sufficient but also a necessary condition for RSB interfaces to be present.

\subsection{Sufficient condition}
\label{subsec:sufficient}

In what follows we separate the discussion into two parts. We first discuss the sufficient condition, which was derived in~\cite{NS22} as Theorem 8.2.

{\bf Theorem 4.1~(Newman-Stein~\cite{NS22}.} If a GSP $\sigma$ chosen from $\kappa_J$ has a positive fraction of edges with space-filling critical droplets, then $\sigma$ will have an RSB interface with one or more other GSP's in~$\kappa_J$.

\medskip

We reproduce the proof from~\cite{NS22} below.

\medskip

{\bf Proof.}  In each finite volume~$\Lambda_L$, choose an arbitrary~edge uniformly at random within~${\mathbb E}_L$ (the edge set restricted to $\Lambda_L$) and consider the excited state~$\tau_L$ generated by flipping its critical droplet (with $J$ remaining fixed).  

By assumption the procedure defined above has a positive probability of generating a positive-density critical droplet, in which case the size of the interface boundary between $\tau_L$ and~$\sigma_L$ scales as $L^d$.  By the usual compactness arguments the set of interfaces between the $\tau_L$'s and $\sigma_L$'s will converge to limiting space-filling interfaces between $\sigma$ and $\tau$, the infinite-volume spin configurations to which $\sigma_L$ and $\tau_L$ converge along one or more subsequences of $\Lambda_L$'s.  By construction the energy of the interface in any volume is twice the flexibility of the chosen edge and must decrease with $L$, so in the infinite-volume limit the energy of the generated interface between $\tau$ and $\sigma$ remains~$O(1)$ in any finite-volume subset of ${\mathbb Z}^d$.

Using this procedure, consider one such edge~$b_1$ chosen in $\mathbb{E}_L$ which has a SFCD in $\sigma_L$.  By definition the critical droplet is the lowest-energy droplet generated by changing an edge's coupling value past its critical value.  Then condition~(\ref{eq:gs}) is satisfied in~$\tau_L$ for all closed contours or surfaces {\it except\/} those passing through $b_1$.  Consider next a fixed cube (a ``window'') centered at the origin whose edge $w$ satisfies $1\ll w\ll L$. Because $b_1$ is chosen uniformly at random within $\Lambda_L$,  it will move outside any fixed window with probability approaching one as $L\to\infty$; therefore~(\ref{eq:gs}) will be satisfied within any fixed window for $\tau$ itself.  Consequently $\tau$ is also an infinite-volume GSP of the Hamiltonian~(\ref{eq:EA}) with a positive-density low-energy interface with $\sigma$. $\diamond$

\subsection{Necessary condition}
\label{subsec:necessary}

In~\cite{NS22} it was shown that a necessary condition for the existence of RSB interfaces was the presence of at least one of two kinds of edges. The first of these consists of edges having SFCD's, and the second includes edges without SFCD's but which lie in the critical droplet boundary of a positive density (in $\mathbb{E}^d$) of other edges. In what follows we show that the second kind of edge is not needed and the presence of SFCD's is by itself a necessary condition. Before we do that, we need to review some earlier results.  To do this, we will need to use the concept of a metastate; an extensive introduction and review can be found in~\cite{NRS23b}. Here we simply note that a metastate is a probability measure on the thermodynamic states of the system; two different constructions can be found in~\cite{AW90} and~\cite{NS96c}.
Without reference to various constructions, a metastate satisfies three properties: first, it is supported solely on the thermodynamic states of a given Hamiltonian generated through an infinite sequence of volumes with prespecified boundary conditions (such as periodic, free, or fixed, for example). Second, it satisfies the property of coupling covariance, meaning that the set of thermodynamic states in the support of the metastate doesn't change when any finite set of couplings are varied.  That is, correlations in the thermodynamic states may change, but every thermodynamic state in the metastate is mapped continuously to a new one as the couplings vary; no thermodynamic states flow into or out of the metastate under a finite change in couplings. And third, the metastate satisfies translation covariance, that is, a uniform lattice shift does not affect the metastate properties.

Using the properties of metastates, Arguin {\it et al.\/}~\cite{ANSW14} proved the following result (which we state here somewhat informally) for the EA Ising model:

\smallskip

{\bf Theorem 4.2~\cite{ANSW14}.} Suppose an edge correlation function $\langle\sigma_x\sigma_y\rangle$ differs with positive probability in two distinct metastates $\kappa_1$ and $\kappa_2$. Choose a thermodynamic state $\Gamma_1$ from the support of $\kappa_1$ and similarly choose a thermodynamic state $\Gamma_2$ from $\kappa_2$, and let $F_L(\Gamma_1,\Gamma_2)$ denote the free energy difference between $\Gamma_1$ and $\Gamma_2$ within the restricted volume $\Lambda_L\in{\mathbb Z}^d$ .  Then there is a constant $c>0$ such that the variance of $F_L(\Gamma,\Gamma')$ with respect to varying the couplings inside~$\Lambda_L$ satisfies
\begin{equation}
\label{eq:flucs}
{\rm Var}\Big(F_L(\Gamma,\Gamma')\Big)\ge c\vert\Lambda_L\vert\, .
\end{equation}

In~\cite{NS24a,NS24b} the authors extended these ideas to a new kind of metastate called the {\it restricted metastate\/}.  The idea behind restricted metastates is to start with a conventional metastate, say that constructed using an infinite sequence of volumes with periodic boundary conditions (call it $\kappa_J$). Next, choose a pure state (call it $\omega$) randomly from $\kappa_J$, and then retain only those pure states in $\kappa_J$ whose edge overlap falls within a narrow prespecified range. The edge overlap between two Gibbs states $\alpha$ and $\alpha'$ is defined to be
\begin{equation}
\label{eq:overlap}
q^{(e)}_{\alpha\alpha'}=\lim_{L\to\infty}\frac{1}{d\vert\Lambda_L\vert}\sum_{\langle xy\rangle\in E_L}\langle\sigma_x\sigma_y\rangle_\alpha\langle\sigma_x\sigma_y\rangle_{\alpha'}\, .
\end{equation}
where $E_L$ denotes the edge set within $\Lambda_L$.
This will generate a nontrivial metastate if $\kappa_J$ contains multiple ``incongruent'' pure states as predicted by RSB, i.e., pairs of pure states whose edge overlap is strictly smaller than their self-overlap.  By choosing different prespecified overlaps, one can construct different restricted metastates that satisfy the conditions of Theorem~4.2, leading to the conclusion that the variance of free energy fluctuations increases linearly with the volume considered.

However, this can be done (so far) only at positive temperature because of the requirement of coupling covariance.  It was shown in~\cite{NS24b} (Lemma 4.1) that at positive temperature $q^{(e)}_{\alpha\alpha'}$ is invariant with respect to a finite change in couplings.  However, it is not necessarily the case that this is true for ground states, because of the possibility of the existence of SFCD's.  But it is also clear that if SFCD's do {\it not\/} exist, then any finite change in couplings can affect only a zero density of edge correlations $\sigma_x\sigma_y$ (with $x$ and $y$ nearest neighbors) in either $\alpha$ or $\alpha'$, now understood to refer to infinite-volume ground states. In this case $q^{(e)}_{\alpha\alpha'}$ again remains invariant under any finite change in couplings, coupling covariance is satisfied, and Theorem 4.2 can now be applied. 

Now if RSB interfaces exist, then there must be ground states in the support of $\kappa_J$ which are mutually incongruent.  Moreover, the magnitude of the energy of an interface (as measured from either $\alpha$ or $\alpha'$) in $\Lambda_L$ equals half the energy difference between $\alpha$ and $\alpha'$ inside~$\Lambda_L$.  But by~(\ref{eq:flucs}) the interface energy between $\alpha$ and $\alpha'$ --- or any other pair of ground states chosen from $\kappa_J$ --- scales with $L$ (typically as $L^{d/2}$); see also Proposition~6.1 of~\cite{AW90}. The conclusion is that no pair of ground states in the support of $\kappa_J$ can differ by an RSB interface if SFCD's exist. We have therefore proved the main new result of this paper:

\medskip

{\bf Theorem 4.3.} Suppose that ground states in the support of the PBC~metastate $\kappa_J$ have no edges with SFCD's. Then RSB interfaces between two ground states are absent in the metastate.

\medskip

Following the discussion in Sect.~12 of~\cite{NS24b} we also have the following corollary:

\medskip

{\bf Corollary 4.4.} If ground states in the support of the two-dimensional zero-temperature PBC~metastate $\kappa_J$ have no edges with SFCD's, then the metastate is supported on a single
pair of spin-reversed ground states.

\section{Discussion}
\label{sec:disc}

Replica symmetry breaking predicts that there exist space-filling, low-energy interfaces between ground states in three and higher dimensions. We have shown that this
prediction is equivalent to the presence of space-filling critical droplets for a positive density of edges in $\mathbb{E}^d$ in a typical ground state; that is, the presence of SFCD's is both a necessary and sufficient condition for the appearance of RSB interfaces.  A stronger conclusion can be drawn in two dimensions, where ground state multiplicity relies on SFCD's: if they're absent, the zero-temperature periodic boundary condition metastate $\kappa_J$ is supported on a single pair of spin-reversed ground states.

Where does this leave the other three scenarios appearing in Table 1? Like RSB, the chaotic pairs scenario also predicts the appearance of multiple incongruent ground states separated by space-filling interfaces, but unlike RSB the interface energy in chaotic pairs scales with~$L$.  To address this scenario, we require the following quantities, introduced in~\cite{NS22}. Let $K^*(b,\sigma)$ denote the number of edges in ${\mathbb E}^d$ whose critical droplet boundaries in ground state~$\sigma$ pass through the edge~$b$. Then for $k = 1,2,3,\ldots$ define $P(k, \sigma)$ to be the fraction of edges~$b\in{\mathbb E}^d$ such that $K^*(b,\sigma) = k$, and let 
\begin{equation}
\label{eq:avg}
E_\sigma [K^*] = \sum_{k=1}^\infty k\ P(k, \sigma)\, .  
\end{equation}
That is, $E_\sigma[K^*]$ is the average number of edges whose critical droplet boundaries a typical edge belongs to in the GSP~$\sigma$.  Using results from this paper and~\cite{NS22}, we conclude that if SFCD's are absent and (a positive fraction of) ground states in $\kappa_J$ are characterized by $E_\sigma[K^*]=\infty$, then the chaotic pairs scenario should hold. 

It follows that neither RSB nor chaotic pairs will hold if $E_\sigma[K^*]<\infty$, which follows if $P(k,\sigma)$ falls off faster than $k^{-(2+\epsilon)}$ for any $\epsilon>0$ as $k\to\infty$. If this is the case then $\kappa_J$ is supported on a single pair of spin-reversed ground states and either scaling-droplet or TNT should hold.

\smallskip

{\it Note added in proof.\/} After this paper went to press, we were made aware of a new study~\cite{VMS24} which suggests a picture where in low dimensions there exists a crossover lengthscale below which RSB-like and droplet-scaling-like excitations coexist (but on different lengthscales) and above which only droplet-scaling-like excitations survive.

\smallskip

{\it Acknowledgments.} The authors thank two anonymous reviewers for their comments on an earlier version which helped to clarify parts of this paper and M.A. Moore for useful discussions and bringing our attention to Ref.~\cite{VMS24} .

\section*{Conflict of Interest Statement}

The authors declare that the research was conducted in the absence of any commercial or financial relationships that could be construed as a potential conflict of interest.

\section*{Author Contributions}

Both authors contributed equally to all aspects of this work.

\section*{Funding}

The authors declare that no financial support was received for the research, authorship, and/or publication of this article.

\bibliography{refs.bib}

\small\def\em{\it} \newcommand{\noopsort}[1]{} \newcommand{\printfirst}[2]{#1}
  \newcommand{\singleletter}[1]{#1} \newcommand{\switchargs}[2]{#2#1}
\begin{thebibliography}{38}%
\makeatletter
\providecommand \@ifxundefined [1]{%
 \@ifx{#1\undefined}
}%
\providecommand \@ifnum [1]{%
 \ifnum #1\expandafter \@firstoftwo
 \else \expandafter \@secondoftwo
 \fi
}%
\providecommand \@ifx [1]{%
 \ifx #1\expandafter \@firstoftwo
 \else \expandafter \@secondoftwo
 \fi
}%
\providecommand \natexlab [1]{#1}%
\providecommand \enquote  [1]{``#1''}%
\providecommand \bibnamefont  [1]{#1}%
\providecommand \bibfnamefont [1]{#1}%
\providecommand \citenamefont [1]{#1}%
\providecommand \href@noop [0]{\@secondoftwo}%
\providecommand \href [0]{\begingroup \@sanitize@url \@href}%
\providecommand \@href[1]{\@@startlink{#1}\@@href}%
\providecommand \@@href[1]{\endgroup#1\@@endlink}%
\providecommand \@sanitize@url [0]{\catcode `\\12\catcode `\$12\catcode
  `\&12\catcode `\#12\catcode `\^12\catcode `\_12\catcode `\%12\relax}%
\providecommand \@@startlink[1]{}%
\providecommand \@@endlink[0]{}%
\providecommand \url  [0]{\begingroup\@sanitize@url \@url }%
\providecommand \@url [1]{\endgroup\@href {#1}{\urlprefix }}%
\providecommand \urlprefix  [0]{URL }%
\providecommand \Eprint [0]{\href }%
\providecommand \doibase [0]{http://dx.doi.org/}%
\providecommand \selectlanguage [0]{\@gobble}%
\providecommand \bibinfo  [0]{\@secondoftwo}%
\providecommand \bibfield  [0]{\@secondoftwo}%
\providecommand \translation [1]{[#1]}%
\providecommand \BibitemOpen [0]{}%
\providecommand \bibitemStop [0]{}%
\providecommand \bibitemNoStop [0]{.\EOS\space}%
\providecommand \EOS [0]{\spacefactor3000\relax}%
\providecommand \BibitemShut  [1]{\csname bibitem#1\endcsname}%
\let\auto@bib@innerbib\@empty
\bibitem [{\citenamefont {Edwards}\ and\ \citenamefont
  {Anderson}(1975)}]{EA75}%
  \BibitemOpen
  \bibfield  {author} {\bibinfo {author} {\bibfnamefont {S.}~\bibnamefont
  {Edwards}}\ and\ \bibinfo {author} {\bibfnamefont {P.~W.}\ \bibnamefont
  {Anderson}},\ }\href@noop {} {\bibfield  {journal} {\bibinfo  {journal} {J.
  Phys. F}\ }\textbf {\bibinfo {volume} {5}},\ \bibinfo {pages} {965} (\bibinfo
  {year} {1975})}\BibitemShut {NoStop}%
\bibitem [{\citenamefont {Parisi}(1979)}]{Parisi79}%
  \BibitemOpen
  \bibfield  {author} {\bibinfo {author} {\bibfnamefont {G.}~\bibnamefont
  {Parisi}},\ }\href@noop {} {\bibfield  {journal} {\bibinfo  {journal} {Phys.
  Rev. Lett.}\ }\textbf {\bibinfo {volume} {43}},\ \bibinfo {pages} {1754}
  (\bibinfo {year} {1979})}\BibitemShut {NoStop}%
\bibitem [{\citenamefont {Parisi}(1983)}]{Parisi83}%
  \BibitemOpen
  \bibfield  {author} {\bibinfo {author} {\bibfnamefont {G.}~\bibnamefont
  {Parisi}},\ }\href@noop {} {\bibfield  {journal} {\bibinfo  {journal} {Phys.
  Rev. Lett.}\ }\textbf {\bibinfo {volume} {50}},\ \bibinfo {pages} {1946}
  (\bibinfo {year} {1983})}\BibitemShut {NoStop}%
\bibitem [{\citenamefont {M\'ezard}\ \emph
  {et~al.}(1984{\natexlab{a}})\citenamefont {M\'ezard}, \citenamefont {Parisi},
  \citenamefont {Sourlas}, \citenamefont {Toulouse},\ and\ \citenamefont
  {Virasoro}}]{MPSTV84a}%
  \BibitemOpen
  \bibfield  {author} {\bibinfo {author} {\bibfnamefont {M.}~\bibnamefont
  {M\'ezard}}, \bibinfo {author} {\bibfnamefont {G.}~\bibnamefont {Parisi}},
  \bibinfo {author} {\bibfnamefont {N.}~\bibnamefont {Sourlas}}, \bibinfo
  {author} {\bibfnamefont {G.}~\bibnamefont {Toulouse}}, \ and\ \bibinfo
  {author} {\bibfnamefont {M.}~\bibnamefont {Virasoro}},\ }\href@noop {}
  {\bibfield  {journal} {\bibinfo  {journal} {Phys. Rev. Lett.}\ }\textbf
  {\bibinfo {volume} {52}},\ \bibinfo {pages} {1156} (\bibinfo {year}
  {1984}{\natexlab{a}})}\BibitemShut {NoStop}%
\bibitem [{\citenamefont {M\'ezard}\ \emph
  {et~al.}(1984{\natexlab{b}})\citenamefont {M\'ezard}, \citenamefont {Parisi},
  \citenamefont {Sourlas}, \citenamefont {Toulouse},\ and\ \citenamefont
  {Virasoro}}]{MPSTV84b}%
  \BibitemOpen
  \bibfield  {author} {\bibinfo {author} {\bibfnamefont {M.}~\bibnamefont
  {M\'ezard}}, \bibinfo {author} {\bibfnamefont {G.}~\bibnamefont {Parisi}},
  \bibinfo {author} {\bibfnamefont {N.}~\bibnamefont {Sourlas}}, \bibinfo
  {author} {\bibfnamefont {G.}~\bibnamefont {Toulouse}}, \ and\ \bibinfo
  {author} {\bibfnamefont {M.}~\bibnamefont {Virasoro}},\ }\href@noop {}
  {\bibfield  {journal} {\bibinfo  {journal} {J. Phys. (Paris)}\ }\textbf
  {\bibinfo {volume} {45}},\ \bibinfo {pages} {843} (\bibinfo {year}
  {1984}{\natexlab{b}})}\BibitemShut {NoStop}%
\bibitem [{\citenamefont {M\'ezard}\ \emph {et~al.}(1987)\citenamefont
  {M\'ezard}, \citenamefont {Parisi},\ and\ \citenamefont {Virasoro}}]{MPV87}%
  \BibitemOpen
  \bibinfo {editor} {\bibfnamefont {M.}~\bibnamefont {M\'ezard}}, \bibinfo
  {editor} {\bibfnamefont {G.}~\bibnamefont {Parisi}}, \ and\ \bibinfo {editor}
  {\bibfnamefont {M.~A.}\ \bibnamefont {Virasoro}},\ eds.,\ \href@noop {}
  {\emph {\bibinfo {title} {Spin Glass Theory and Beyond}}}\ (\bibinfo
  {publisher} {World Scientific},\ \bibinfo {address} {Singapore},\ \bibinfo
  {year} {1987})\BibitemShut {NoStop}%
\bibitem [{\citenamefont {Marinari}\ \emph {et~al.}(1996)\citenamefont
  {Marinari}, \citenamefont {Parisi}, \citenamefont {Ruiz-Lorenzo},\ and\
  \citenamefont {Ritort}}]{MPRR96}%
  \BibitemOpen
  \bibfield  {author} {\bibinfo {author} {\bibfnamefont {E.}~\bibnamefont
  {Marinari}}, \bibinfo {author} {\bibfnamefont {G.}~\bibnamefont {Parisi}},
  \bibinfo {author} {\bibfnamefont {J.~J.}\ \bibnamefont {Ruiz-Lorenzo}}, \
  and\ \bibinfo {author} {\bibfnamefont {F.}~\bibnamefont {Ritort}},\
  }\href@noop {} {\bibfield  {journal} {\bibinfo  {journal} {Phys. Rev. Lett.}\
  }\textbf {\bibinfo {volume} {76}},\ \bibinfo {pages} {843} (\bibinfo {year}
  {1996})}\BibitemShut {NoStop}%
\bibitem [{\citenamefont {Marinari}\ \emph {et~al.}(2000)\citenamefont
  {Marinari}, \citenamefont {Parisi}, \citenamefont {Ricci-Tersenghi},
  \citenamefont {Ruiz-Lorenzo},\ and\ \citenamefont {Zuliani}}]{MPRRZ00}%
  \BibitemOpen
  \bibfield  {author} {\bibinfo {author} {\bibfnamefont {E.}~\bibnamefont
  {Marinari}}, \bibinfo {author} {\bibfnamefont {G.}~\bibnamefont {Parisi}},
  \bibinfo {author} {\bibfnamefont {F.}~\bibnamefont {Ricci-Tersenghi}},
  \bibinfo {author} {\bibfnamefont {J.~J.}\ \bibnamefont {Ruiz-Lorenzo}}, \
  and\ \bibinfo {author} {\bibfnamefont {F.}~\bibnamefont {Zuliani}},\
  }\href@noop {} {\bibfield  {journal} {\bibinfo  {journal} {J. Stat. Phys.}\
  }\textbf {\bibinfo {volume} {98}},\ \bibinfo {pages} {973} (\bibinfo {year}
  {2000})}\BibitemShut {NoStop}%
\bibitem [{\citenamefont {Newman}\ and\ \citenamefont {Stein}(2002)}]{NS02}%
  \BibitemOpen
  \bibfield  {author} {\bibinfo {author} {\bibfnamefont {C.~M.}\ \bibnamefont
  {Newman}}\ and\ \bibinfo {author} {\bibfnamefont {D.~L.}\ \bibnamefont
  {Stein}},\ }\href@noop {} {\bibfield  {journal} {\bibinfo  {journal} {J.
  Stat. Phys.}\ }\textbf {\bibinfo {volume} {106}},\ \bibinfo {pages} {213}
  (\bibinfo {year} {2002})}\BibitemShut {NoStop}%
\bibitem [{\citenamefont {Newman}\ and\ \citenamefont
  {Stein}(2003{\natexlab{a}})}]{NS03b}%
  \BibitemOpen
  \bibfield  {author} {\bibinfo {author} {\bibfnamefont {C.~M.}\ \bibnamefont
  {Newman}}\ and\ \bibinfo {author} {\bibfnamefont {D.~L.}\ \bibnamefont
  {Stein}},\ }\href@noop {} {\bibfield  {journal} {\bibinfo  {journal} {J.
  Phys.: Cond. Mat.}\ }\textbf {\bibinfo {volume} {15}},\ \bibinfo {pages}
  {R1319 } (\bibinfo {year} {2003}{\natexlab{a}})}\BibitemShut {NoStop}%
\bibitem [{\citenamefont {Read}(2014)}]{Read14}%
  \BibitemOpen
  \bibfield  {author} {\bibinfo {author} {\bibfnamefont {N.}~\bibnamefont
  {Read}},\ }\href@noop {} {\bibfield  {journal} {\bibinfo  {journal} {Phys.
  Rev. E}\ }\textbf {\bibinfo {volume} {90}},\ \bibinfo {pages} {032142}
  (\bibinfo {year} {2014})}\BibitemShut {NoStop}%
\bibitem [{\citenamefont {Newman}\ \emph {et~al.}(2023)\citenamefont {Newman},
  \citenamefont {Read},\ and\ \citenamefont {Stein}}]{NRS23b}%
  \BibitemOpen
  \bibfield  {author} {\bibinfo {author} {\bibfnamefont {C.~M.}\ \bibnamefont
  {Newman}}, \bibinfo {author} {\bibfnamefont {N.}~\bibnamefont {Read}}, \ and\
  \bibinfo {author} {\bibfnamefont {D.~L.}\ \bibnamefont {Stein}},\ }in\
  \href@noop {} {\emph {\bibinfo {booktitle} {Spin Glass Theory and Far Beyond:
  Replica Symmetry Breaking After 40 Years}}},\ \bibinfo {editor} {edited by\
  \bibinfo {editor} {\bibfnamefont {P.}~\bibnamefont {Charbonneau}}, \bibinfo
  {editor} {\bibfnamefont {E.}~\bibnamefont {Marinari}}, \bibinfo {editor}
  {\bibfnamefont {M.}~\bibnamefont {M{\'e}zard}}, \bibinfo {editor}
  {\bibfnamefont {G.}~\bibnamefont {Parisi}}, \bibinfo {editor} {\bibfnamefont
  {F.}~\bibnamefont {Ricci-Tersenghi}}, \bibinfo {editor} {\bibfnamefont
  {G.}~\bibnamefont {Sicuro}}, \ and\ \bibinfo {editor} {\bibfnamefont
  {F.}~\bibnamefont {Zamponi}}}\ (\bibinfo  {publisher} {Singapore: World
  Scientific},\ \bibinfo {year} {2023})\ pp.\ \bibinfo {pages}
  {697--718}\BibitemShut {NoStop}%
\bibitem [{\citenamefont {McMillan}(1984)}]{Mac84}%
  \BibitemOpen
  \bibfield  {author} {\bibinfo {author} {\bibfnamefont {W.~L.}\ \bibnamefont
  {McMillan}},\ }\href@noop {} {\bibfield  {journal} {\bibinfo  {journal} {J.
  Phys. C}\ }\textbf {\bibinfo {volume} {17}},\ \bibinfo {pages} {3179}
  (\bibinfo {year} {1984})}\BibitemShut {NoStop}%
\bibitem [{\citenamefont {Bray}\ and\ \citenamefont {Moore}(1985)}]{BM85}%
  \BibitemOpen
  \bibfield  {author} {\bibinfo {author} {\bibfnamefont {A.~J.}\ \bibnamefont
  {Bray}}\ and\ \bibinfo {author} {\bibfnamefont {M.~A.}\ \bibnamefont
  {Moore}},\ }\href@noop {} {\bibfield  {journal} {\bibinfo  {journal} {Phys.
  Rev. B}\ }\textbf {\bibinfo {volume} {31}},\ \bibinfo {pages} {631} (\bibinfo
  {year} {1985})}\BibitemShut {NoStop}%
\bibitem [{\citenamefont {Bray}\ and\ \citenamefont {Moore}(1987)}]{BM87}%
  \BibitemOpen
  \bibfield  {author} {\bibinfo {author} {\bibfnamefont {A.~J.}\ \bibnamefont
  {Bray}}\ and\ \bibinfo {author} {\bibfnamefont {M.~A.}\ \bibnamefont
  {Moore}},\ }\href@noop {} {\bibfield  {journal} {\bibinfo  {journal} {Phys.
  Rev. Lett.}\ }\textbf {\bibinfo {volume} {58}},\ \bibinfo {pages} {57}
  (\bibinfo {year} {1987})}\BibitemShut {NoStop}%
\bibitem [{\citenamefont {Fisher}\ and\ \citenamefont {Huse}(1986)}]{FH86}%
  \BibitemOpen
  \bibfield  {author} {\bibinfo {author} {\bibfnamefont {D.~S.}\ \bibnamefont
  {Fisher}}\ and\ \bibinfo {author} {\bibfnamefont {D.~A.}\ \bibnamefont
  {Huse}},\ }\href@noop {} {\bibfield  {journal} {\bibinfo  {journal} {Phys.
  Rev. Lett.}\ }\textbf {\bibinfo {volume} {56}},\ \bibinfo {pages} {1601}
  (\bibinfo {year} {1986})}\BibitemShut {NoStop}%
\bibitem [{\citenamefont {Fisher}\ and\ \citenamefont {Huse}(1988)}]{FH88b}%
  \BibitemOpen
  \bibfield  {author} {\bibinfo {author} {\bibfnamefont {D.~S.}\ \bibnamefont
  {Fisher}}\ and\ \bibinfo {author} {\bibfnamefont {D.~A.}\ \bibnamefont
  {Huse}},\ }\href@noop {} {\bibfield  {journal} {\bibinfo  {journal} {Phys.
  Rev. B}\ }\textbf {\bibinfo {volume} {38}},\ \bibinfo {pages} {386} (\bibinfo
  {year} {1988})}\BibitemShut {NoStop}%
\bibitem [{\citenamefont {Marinari}\ and\ \citenamefont {Parisi}(2000)}]{MP00}%
  \BibitemOpen
  \bibfield  {author} {\bibinfo {author} {\bibfnamefont {E.}~\bibnamefont
  {Marinari}}\ and\ \bibinfo {author} {\bibfnamefont {G.}~\bibnamefont
  {Parisi}},\ }\href@noop {} {\bibfield  {journal} {\bibinfo  {journal} {Phys.
  Rev. B}\ }\textbf {\bibinfo {volume} {62}},\ \bibinfo {pages} {11677}
  (\bibinfo {year} {2000})}\BibitemShut {NoStop}%
\bibitem [{\citenamefont {Palassini}\ and\ \citenamefont {Young}(2000)}]{PY00}%
  \BibitemOpen
  \bibfield  {author} {\bibinfo {author} {\bibfnamefont {M.}~\bibnamefont
  {Palassini}}\ and\ \bibinfo {author} {\bibfnamefont {A.~P.}\ \bibnamefont
  {Young}},\ }\href@noop {} {\bibfield  {journal} {\bibinfo  {journal} {Phys.
  Rev. Lett.}\ }\textbf {\bibinfo {volume} {85}},\ \bibinfo {pages} {3017}
  (\bibinfo {year} {2000})}\BibitemShut {NoStop}%
\bibitem [{\citenamefont {Newman}\ and\ \citenamefont {Stein}(1996)}]{NS96c}%
  \BibitemOpen
  \bibfield  {author} {\bibinfo {author} {\bibfnamefont {C.~M.}\ \bibnamefont
  {Newman}}\ and\ \bibinfo {author} {\bibfnamefont {D.~L.}\ \bibnamefont
  {Stein}},\ }\href@noop {} {\bibfield  {journal} {\bibinfo  {journal} {Phys.
  Rev. Lett.}\ }\textbf {\bibinfo {volume} {76}},\ \bibinfo {pages} {4821}
  (\bibinfo {year} {1996})}\BibitemShut {NoStop}%
\bibitem [{\citenamefont {Newman}\ and\ \citenamefont {Stein}(1997)}]{NS97}%
  \BibitemOpen
  \bibfield  {author} {\bibinfo {author} {\bibfnamefont {C.~M.}\ \bibnamefont
  {Newman}}\ and\ \bibinfo {author} {\bibfnamefont {D.~L.}\ \bibnamefont
  {Stein}},\ }\href@noop {} {\bibfield  {journal} {\bibinfo  {journal} {Phys.
  Rev. E}\ }\textbf {\bibinfo {volume} {55}},\ \bibinfo {pages} {5194}
  (\bibinfo {year} {1997})}\BibitemShut {NoStop}%
\bibitem [{\citenamefont {Newman}\ and\ \citenamefont
  {Stein}(1998)}]{NSBerlin}%
  \BibitemOpen
  \bibfield  {author} {\bibinfo {author} {\bibfnamefont {C.~M.}\ \bibnamefont
  {Newman}}\ and\ \bibinfo {author} {\bibfnamefont {D.~L.}\ \bibnamefont
  {Stein}},\ }in\ \href@noop {} {\emph {\bibinfo {booktitle} {Mathematics of
  Spin Glasses and Neural Networks}}},\ \bibinfo {editor} {edited by\ \bibinfo
  {editor} {\bibfnamefont {A.}~\bibnamefont {Bovier}}\ and\ \bibinfo {editor}
  {\bibfnamefont {P.}~\bibnamefont {Picco}}}\ (\bibinfo  {publisher}
  {Birkhauser},\ \bibinfo {address} {Boston},\ \bibinfo {year} {1998})\ pp.\
  \bibinfo {pages} {243--287}\BibitemShut {NoStop}%
\bibitem [{\citenamefont {Newman}\ and\ \citenamefont
  {Stein}(2003{\natexlab{b}})}]{NS03a}%
  \BibitemOpen
  \bibfield  {author} {\bibinfo {author} {\bibfnamefont {C.~M.}\ \bibnamefont
  {Newman}}\ and\ \bibinfo {author} {\bibfnamefont {D.~L.}\ \bibnamefont
  {Stein}},\ }\href@noop {} {\bibfield  {journal} {\bibinfo  {journal} {Ann.
  Henri Poinca{r\'e,} Suppl. 1}\ }\textbf {\bibinfo {volume} {4}},\ \bibinfo
  {pages} {S497 } (\bibinfo {year} {2003}{\natexlab{b}})}\BibitemShut {NoStop}%
\bibitem [{\citenamefont {Newman}\ and\ \citenamefont {Stein}(2022)}]{NS22}%
  \BibitemOpen
  \bibfield  {author} {\bibinfo {author} {\bibfnamefont {C.~M.}\ \bibnamefont
  {Newman}}\ and\ \bibinfo {author} {\bibfnamefont {D.~L.}\ \bibnamefont
  {Stein}},\ }\href@noop {} {\bibfield  {journal} {\bibinfo  {journal} {Phys.
  Rev. E}\ }\textbf {\bibinfo {volume} {105}},\ \bibinfo {pages} {044132}
  (\bibinfo {year} {2022})}\BibitemShut {NoStop}%
\bibitem [{\citenamefont {Newman}\ and\ \citenamefont
  {Stein}(2001{\natexlab{a}})}]{NS01c}%
  \BibitemOpen
  \bibfield  {author} {\bibinfo {author} {\bibfnamefont {C.~M.}\ \bibnamefont
  {Newman}}\ and\ \bibinfo {author} {\bibfnamefont {D.~L.}\ \bibnamefont
  {Stein}},\ }\href@noop {} {\bibfield  {journal} {\bibinfo  {journal} {Phys.
  Rev. Lett.}\ }\textbf {\bibinfo {volume} {87}},\ \bibinfo {pages} {077201}
  (\bibinfo {year} {2001}{\natexlab{a}})}\BibitemShut {NoStop}%
\bibitem [{\citenamefont {Arguin}\ \emph {et~al.}(2019)\citenamefont {Arguin},
  \citenamefont {Newman},\ and\ \citenamefont {Stein}}]{ANS19}%
  \BibitemOpen
  \bibfield  {author} {\bibinfo {author} {\bibfnamefont {L.-P.}\ \bibnamefont
  {Arguin}}, \bibinfo {author} {\bibfnamefont {C.~M.}\ \bibnamefont {Newman}},
  \ and\ \bibinfo {author} {\bibfnamefont {D.~L.}\ \bibnamefont {Stein}},\
  }\href@noop {} {\bibfield  {journal} {\bibinfo  {journal} {Commun. Math.
  Phys.}\ }\textbf {\bibinfo {volume} {367}},\ \bibinfo {pages} {1019}
  (\bibinfo {year} {2019})}\BibitemShut {NoStop}%
\bibitem [{\citenamefont {{de Almeida}}\ and\ \citenamefont
  {Thouless}(1978)}]{AT78}%
  \BibitemOpen
  \bibfield  {author} {\bibinfo {author} {\bibfnamefont {J.~R.~L.}\
  \bibnamefont {{de Almeida}}}\ and\ \bibinfo {author} {\bibfnamefont {D.~J.}\
  \bibnamefont {Thouless}},\ }\href@noop {} {\bibfield  {journal} {\bibinfo
  {journal} {J. Phys. A}\ }\textbf {\bibinfo {volume} {11}},\ \bibinfo {pages}
  {983} (\bibinfo {year} {1978})}\BibitemShut {NoStop}%
\bibitem [{\citenamefont {Huse}\ and\ \citenamefont {Fisher}(1987)}]{HF87}%
  \BibitemOpen
  \bibfield  {author} {\bibinfo {author} {\bibfnamefont {D.~A.}\ \bibnamefont
  {Huse}}\ and\ \bibinfo {author} {\bibfnamefont {D.~S.}\ \bibnamefont
  {Fisher}},\ }\href@noop {} {\bibfield  {journal} {\bibinfo  {journal} {J.
  Phys. A}\ }\textbf {\bibinfo {volume} {20}},\ \bibinfo {pages} {L997}
  (\bibinfo {year} {1987})}\BibitemShut {NoStop}%
\bibitem [{\citenamefont {Fisher}\ and\ \citenamefont {Huse}(1987)}]{FH87}%
  \BibitemOpen
  \bibfield  {author} {\bibinfo {author} {\bibfnamefont {D.~S.}\ \bibnamefont
  {Fisher}}\ and\ \bibinfo {author} {\bibfnamefont {D.~A.}\ \bibnamefont
  {Huse}},\ }\href@noop {} {\bibfield  {journal} {\bibinfo  {journal} {J. Phys.
  A}\ }\textbf {\bibinfo {volume} {20}},\ \bibinfo {pages} {L1005} (\bibinfo
  {year} {1987})}\BibitemShut {NoStop}%
\bibitem [{\citenamefont {Newman}\ and\ \citenamefont {Stein}(2000)}]{NS2D00}%
  \BibitemOpen
  \bibfield  {author} {\bibinfo {author} {\bibfnamefont {C.~M.}\ \bibnamefont
  {Newman}}\ and\ \bibinfo {author} {\bibfnamefont {D.~L.}\ \bibnamefont
  {Stein}},\ }\href@noop {} {\bibfield  {journal} {\bibinfo  {journal} {Phys.
  Rev. Lett.}\ }\textbf {\bibinfo {volume} {84}},\ \bibinfo {pages} {3966}
  (\bibinfo {year} {2000})}\BibitemShut {NoStop}%
\bibitem [{\citenamefont {Newman}\ and\ \citenamefont
  {Stein}(2001{\natexlab{b}})}]{NS2D01}%
  \BibitemOpen
  \bibfield  {author} {\bibinfo {author} {\bibfnamefont {C.~M.}\ \bibnamefont
  {Newman}}\ and\ \bibinfo {author} {\bibfnamefont {D.~L.}\ \bibnamefont
  {Stein}},\ }\href@noop {} {\bibfield  {journal} {\bibinfo  {journal} {Commun.
  Math. Phys.}\ }\textbf {\bibinfo {volume} {224}},\ \bibinfo {pages} {205}
  (\bibinfo {year} {2001}{\natexlab{b}})}\BibitemShut {NoStop}%
\bibitem [{\citenamefont {Arguin}\ \emph {et~al.}(2021)\citenamefont {Arguin},
  \citenamefont {Newman},\ and\ \citenamefont {Stein}}]{ANS21}%
  \BibitemOpen
  \bibfield  {author} {\bibinfo {author} {\bibfnamefont {L.-P.}\ \bibnamefont
  {Arguin}}, \bibinfo {author} {\bibfnamefont {C.~M.}\ \bibnamefont {Newman}},
  \ and\ \bibinfo {author} {\bibfnamefont {D.~L.}\ \bibnamefont {Stein}},\ }in\
  \href@noop {} {\emph {\bibinfo {booktitle} {In and Out of Equilibrium 3:
  Celebrating Vladas Sidoravicius}}},\ \bibinfo {editor} {edited by\ \bibinfo
  {editor} {\bibfnamefont {M.~E.}\ \bibnamefont {Vares}}, \bibinfo {editor}
  {\bibfnamefont {R.}~\bibnamefont {Fernandez}}, \bibinfo {editor}
  {\bibfnamefont {L.~R.}\ \bibnamefont {Fontes}}, \ and\ \bibinfo {editor}
  {\bibfnamefont {C.~M.}\ \bibnamefont {Newman}}}\ (\bibinfo  {publisher}
  {Birkh{\"a}user},\ \bibinfo {address} {Cham (ZG)},\ \bibinfo {year} {2021})\
  pp.\ \bibinfo {pages} {17--25}\BibitemShut {NoStop}%
\bibitem [{\citenamefont {Arguin}\ \emph {et~al.}(2010)\citenamefont {Arguin},
  \citenamefont {Damron}, \citenamefont {Newman},\ and\ \citenamefont
  {Stein}}]{ADNS10}%
  \BibitemOpen
  \bibfield  {author} {\bibinfo {author} {\bibfnamefont {L.-P.}\ \bibnamefont
  {Arguin}}, \bibinfo {author} {\bibfnamefont {M.}~\bibnamefont {Damron}},
  \bibinfo {author} {\bibfnamefont {C.~M.}\ \bibnamefont {Newman}}, \ and\
  \bibinfo {author} {\bibfnamefont {D.~L.}\ \bibnamefont {Stein}},\ }\href@noop
  {} {\bibfield  {journal} {\bibinfo  {journal} {Commun. Math. Phys.}\ }\textbf
  {\bibinfo {volume} {300}},\ \bibinfo {pages} {641} (\bibinfo {year}
  {2010})}\BibitemShut {NoStop}%
\bibitem [{\citenamefont {Newman}\ and\ \citenamefont
  {Stein}(2024{\natexlab{a}})}]{NS24a}%
  \BibitemOpen
  \bibfield  {author} {\bibinfo {author} {\bibfnamefont {C.~M.}\ \bibnamefont
  {Newman}}\ and\ \bibinfo {author} {\bibfnamefont {D.~L.}\ \bibnamefont
  {Stein}},\ }\href@noop {} {\bibfield  {journal} {\bibinfo  {journal} {J.
  Phys. A : Math. Theor.}\ }\textbf {\bibinfo {volume} {57}},\ \bibinfo {pages}
  {11LT01} (\bibinfo {year} {2024}{\natexlab{a}})}\BibitemShut {NoStop}%
\bibitem [{\citenamefont {Newman}\ and\ \citenamefont
  {Stein}(2024{\natexlab{b}})}]{NS24b}%
  \BibitemOpen
  \bibfield  {author} {\bibinfo {author} {\bibfnamefont {C.~M.}\ \bibnamefont
  {Newman}}\ and\ \bibinfo {author} {\bibfnamefont {D.~L.}\ \bibnamefont
  {Stein}},\ }\href@noop {} {\  (\bibinfo {year} {2024}{\natexlab{b}})},\
  \bibinfo {note} {arXiv:2407.17506}\BibitemShut {NoStop}%
\bibitem [{\citenamefont {Aizenman}\ and\ \citenamefont {Wehr}(1990)}]{AW90}%
  \BibitemOpen
  \bibfield  {author} {\bibinfo {author} {\bibfnamefont {M.}~\bibnamefont
  {Aizenman}}\ and\ \bibinfo {author} {\bibfnamefont {J.}~\bibnamefont
  {Wehr}},\ }\href@noop {} {\bibfield  {journal} {\bibinfo  {journal} {Commun.
  Math. Phys.}\ }\textbf {\bibinfo {volume} {130}},\ \bibinfo {pages} {489}
  (\bibinfo {year} {1990})}\BibitemShut {NoStop}%
\bibitem [{\citenamefont {Arguin}\ \emph {et~al.}(2014)\citenamefont {Arguin},
  \citenamefont {Newman}, \citenamefont {Stein},\ and\ \citenamefont
  {Wehr}}]{ANSW14}%
  \BibitemOpen
  \bibfield  {author} {\bibinfo {author} {\bibfnamefont {L.-P.}\ \bibnamefont
  {Arguin}}, \bibinfo {author} {\bibfnamefont {C.~M.}\ \bibnamefont {Newman}},
  \bibinfo {author} {\bibfnamefont {D.~L.}\ \bibnamefont {Stein}}, \ and\
  \bibinfo {author} {\bibfnamefont {J.}~\bibnamefont {Wehr}},\ }\href@noop {}
  {\bibfield  {journal} {\bibinfo  {journal} {J. Stat. Phys.}\ }\textbf
  {\bibinfo {volume} {156}},\ \bibinfo {pages} {221} (\bibinfo {year}
  {2014})}\BibitemShut {NoStop}%
\bibitem [{\citenamefont {Vedula}\ \emph {et~al.}(2024)\citenamefont {Vedula},
  \citenamefont {Moore},\ and\ \citenamefont {Sharma}}]{VMS24}%
  \BibitemOpen
  \bibfield  {author} {\bibinfo {author} {\bibfnamefont {B.}~\bibnamefont
  {Vedula}}, \bibinfo {author} {\bibfnamefont {M.~A.}\ \bibnamefont {Moore}}, \
  and\ \bibinfo {author} {\bibfnamefont {A.}~\bibnamefont {Sharma}},\
  }\href@noop {} {\  (\bibinfo {year} {2024})},\ \bibinfo {note} {available at
  arXiv:2410.19069}\BibitemShut {NoStop}%
\end{thebibliography}%

\end{document}